\documentclass[sigplan]{acmart}
\renewcommand\footnotetextcopyrightpermission[1]{}

\usepackage[frozencache,cachedir=minted-cache]{minted}

\usepackage{graphicx}
\usepackage{fancybox}
\usepackage{listings}
\usepackage{xcolor}
\usepackage[noabbrev,capitalize]{cleveref}

\definecolor{ballblue}{rgb}{0.13, 0.67, 0.8}
\definecolor{bondiblue}{rgb}{0.0, 0.58, 0.71}
\definecolor{cobalt}{rgb}{0.0, 0.28, 0.67}
\definecolor{copper}{rgb}{0.72, 0.45, 0.2}
\definecolor{darkblue}{rgb}{0.0, 0.0, 0.55}

\newcommand{\tianyin}[1]{{{\color{red} [ty: #1]}}}
\newcommand{\para}[1]{\smallskip\noindent {\bf #1} }
\newcommand{\new}[1]{\textcolor{black}{#1}}
\newcommand{\newjinghao}[1]{\textcolor{black}{#1}}
\newcommand{\comments}[1]{}

\definecolor{darkspringgreen}{rgb}{0.09, 0.45, 0.27}

\lstdefinelanguage{myJava}[]{C}{
  morekeywords={assert},
  morecomment=[f][\color{blue}]{@@},     
  morecomment=[f][\color{red}]-,         
  morecomment=[f][\color{ForestGreen}]+, 
  morecomment=[f][\color{magenta}]{---}, 
  morecomment=[f][\color{magenta}]{+++},
  morecomment=[f][\color{magenta}]{//},
  morecomment=[f][\color{magenta}]{/*},
  morecomment=[f][\color{magenta}]{\ \ \ \ //},
  morecomment=[f][\color{magenta}]{//},
  keywords = {BPF_PROG_TYPE_SECCOMP, SECCOMP_FILTER_FLAG_EXTENDED, SECCOMP_RET_ERRNO, SECCOMP_RET_ALLOW, EPERM},
  keywords = [2]{seccomp, bpf_prog_load, int, if, return, switch, return, case, break, bpf_wait_for_syscall, u64, u8, unsigned, struct, sizeof},
  keywordstyle=\color{purple},
  keywordstyle=[2]\color{blue},
  comment=[l]{//},
  commentstyle=\color{darkspringgreen},
  stringstyle=\color{blue},
  numbers=left,
  basicstyle=\footnotesize\ttfamily,
  numbersep=6pt,
  numberstyle=\tiny\ttfamily,
  breaklines=true,
  escapeinside={(*@}{@*)},
  showstringspaces=false,
  xleftmargin=8pt,
}

\lstdefinelanguage{myGo}[]{}{
  keywords = {Exec, InitSeccomp, finalizeNamespace},
  keywords = [2]{if, return, switch, return, case, break},
  morecomment=[f][\color{magenta}]{//},
  morecomment=[f][\color{magenta}]{\ \ \ \ //},
  morecomment=[f][\color{magenta}]{//},
  keywordstyle=\color{purple},
  keywordstyle=[2]\color{blue},
  comment=[l]{//},
  commentstyle=\color{darkspringgreen},
  stringstyle=\color{blue},
  numbers=left,
  basicstyle=\footnotesize\ttfamily,
  numbersep=6pt,
  numberstyle=\tiny\ttfamily,
  breaklines=true,
  escapeinside={(*@}{@*)},
  showstringspaces=false,
  xleftmargin=8pt,
}

\definecolor{VeryLightGray}{gray}{0.90}

\usepackage{caption}
\usepackage{subcaption}

\newif\ifdraft
\drafttrue

\date{}

\begin{document}

\title{Programmable System Call Security with eBPF}

\author{Jinghao Jia$^1$, YiFei Zhu$^2$, Dan Williams$^3$, Andrea Arcangeli$^4$, Claudio Canella$^5$, Hubertus Franke$^6$,\\
  Tobin Feldman-Fitzthum$^6$, Dimitrios Skarlatos$^7$, Daniel Gruss$^8$, Tianyin Xu$^1$}
\affiliation{%
  \institution{$^1$University of Illinois at Urbana-Champaign, Urbana, IL, USA\\
  $^2$Google, Inc., Sunnyvale, CA, USA \\
  $^3$Virginia Tech, Blacksburg, VA, USA\\
  $^4$Red Hat, Inc., New York, NY, USA\\
  $^5$Amazon Web Services, Graz, Austria\\
  $^6$IBM Research, Yorktown Heights, NY, USA\\
  $^7$Carnegie Mellon University, Pittsburgh, PA, USA\\
  $^8$Graz University of Technology, Graz, Austria}
  \city{}
  \state{}
  \country{}
}

\begin{abstract}

\new{System call filtering is a widely used security mechanism
  for protecting a shared OS kernel
  against untrusted user applications.
However, existing system call filtering techniques either are
  too expensive due to the context switch overhead imposed
  by userspace agents,
  or lack sufficient programmability to express
  advanced policies.
Seccomp, Linux's system call filtering module, is widely used
  by modern container technologies, mobile apps,
  and system management services.
Despite the adoption of the {\it classic} BPF language (cBPF),
  security policies in Seccomp are mostly limited to
  {\it static} allow lists, primarily because cBPF does not support {\it stateful} policies.
Consequently, many essential security features
  cannot be expressed
  precisely and/or require kernel modifications.}

\new{In this paper, we present a programmable system call filtering mechanism, which
  enables more advanced security policies to be expressed by leveraging the {\it extended} BPF language
  (eBPF).
  More specifically, we create a new {\em Seccomp eBPF} program
    type, exposing, modifying or creating new eBPF helper functions to
    safely manage filter state, access kernel and user state, and
    utilize synchronization primitives.  Importantly, our system
    integrates with existing kernel privilege and capability
    mechanisms, enabling unprivileged users to install advanced filters safely.
Our evaluation shows that our eBPF-based filtering can
  enhance existing policies (e.g., reducing the attack surface
    of early execution phase by up to 55.4\% for temporal specialization),
  mitigate real-world vulnerabilities,
  and accelerate filters.}

\end{abstract}

\maketitle
\pagestyle{plain}

\section{Introduction}
\label{sec:intro}

Modern computer systems run a large variety of untrusted applications
  on a trusted operating system (OS) kernel.
These applications interact with the OS kernel through the system call interface.
Hence, system call security
  is a cornerstone for protecting a shared kernel against untrusted user processes.
System call filtering is a widely used system call security mechanism.
The basic idea is to restrict the system calls a given process
  can invoke based on predefined security policies,
  thereby reducing the attack surface.
The filtering is done at the entry point of every system call to
    decide whether to allow or deny each system call.

Early system call filtering techniques such as Janus~\cite{Wagner:1999}
  and Ostia~\cite{Garfinkel:2004} employ
  trusted userspace agents to implement system call security policies.
However, userspace agents incur significant context switch overheads to system call
  filtering, because every system call
  needs to switch to user space for policy checking and then switch back.
Furthermore, the implementations are prone to many common
  traps and pitfalls, such as time-of-check-to-time-of-use (TOCTTOU) based
  race conditions of argument values~\cite{Garfinkel:2003,Bhattacharyya:usec:22,Payer:vee:12}.

Modern system call filter techniques such as Linux's Seccomp (SECure COMPuting)
  run entirely inside the OS kernel,
  without the overhead of additional context switches.
This brings significant performance advantages over userspace agents
  and some resistance to TOCTTOU-based argument races.
Today, Seccomp is widely used to provide
  ``{\it the most important security isolation boundary}~\cite{firecracker}.''
For example, every Android app is isolated using Seccomp~\cite{Lawrence:2017};
\texttt{systemd} uses Seccomp for user process sandboxing~\cite{Corbet:2012};
Google's Sandboxed API project~\cite{sapi} uses Seccomp for sandboxing C/C++ libraries;
Lightweight virtualization technologies
  such as Docker~\cite{Docker_seccomp_doc},
  Google gVisor~\cite{gvisor:profile},
  Amazon Firecraker~\cite{Firecracker_design,firecracker}, LXC/LXD~\cite{lxd}, Rkt~\cite{rtk_seccomp},
  and Kubernetes~\cite{kb8:seccomp} 
  all use Seccomp.

However, a major limitation of Seccomp is the lack of
  sufficient programmability to express advanced security policies.
From an initial mode (known as strict mode) that blocked all system calls
except
  \texttt{\small read()}, \texttt{\small write()},
  \texttt{\small \_exit()}, and \texttt{\small sigreturn()},
Seccomp now (since Linux v3.5) allows custom security policies to be
  written in the {\it classic}
  BPF language (cBPF)~\cite{bpf:vm:lwn} in the filter mode.
Filter mode enables application-specific security policies
  and results in the wide adoptions of Seccomp by different applications.





Unfortunately, the programmability of cBPF is overly lim-
  ited---security policies
  in Seccomp are mostly limited to {\it static} allow lists.
This is primarily because cBPF provides no mechanism to store states
  and hence cBPF filters have to be {\it stateless}.
Furthermore, cBPF provides no interface to
  invoke any other kernel utilities or other BPF programs.
As a result, many desirable and/or essential system call filtering features
  cannot be directly implemented based on Seccomp-cBPF,
  instead requiring significant kernel modifications (a major deployment obstacle).
Section~\ref{sec:usecases} discusses these features in details.
Recognizing the need for more expressive policies, Seccomp recently added a new feature known as
  Notifier~\cite{utrap:brauner:blog}, to support the old idea of userspace agents, which unfortunately shares their limitations
  (performance overhead, TOCTTOU issues, etc.).



In this paper, we present a programmable system call filtering mechanism
  by leveraging the {\it extended} BPF language (eBPF).
Our goal is to enable advanced system call security policies to better
  protect the shared OS kernel, without impairing the system call performance
  or reducing OS security.

Our choice of eBPF is a result of practicality considerations:
  1) eBPF offers basic building blocks
    for the target programmability, including maps for statefulness
    and helper functions for interfacing with the kernel;
  2) like cBPF, eBPF is verified to be safe by the kernel;
  and 3) we can largely reuse existing implementation code in Linux.


Note that na\"ively opening the eBPF interface in Seccomp,
  as early Linux patches~\cite{Dhillon:2018,Hromatka:2018,Zhu:2021},
  is not a solution, because:
  1) basic support is missing (e.g., synchronization primitives for serialization);
  2) existing utilities (e.g., task storage) may not fit the Seccomp model because
    they target privileged contexts only;
and 3) existing features are not safe for Seccomp use cases, e.g., the
  current user memory access feature cannot address TOCTTOU issues.

  To this end,
we create a new Seccomp-eBPF program
  type which is highly programmable for users to
  express advanced system call security policies
  in eBPF filter programs.
Specifically, we expose, modify, and create new eBPF helper functions to
  safely manage filter state, access kernel and user state, as well as
  utilize synchronization primitives. Importantly, our system
  integrates with existing kernel privilege and capability
  mechanisms, enabling unprivileged users to install advanced filters safely
  and preventing privilege escalation.
The security of Seccomp-eBPF is equivalent to the two
  existing kernel components: Seccomp and eBPF.

We implement the new Seccomp-eBPF program type on top of Seccomp in
  the Linux kernel.
We maintain the existing Seccomp interface with tamper protection.
We implement many features required by real-world use cases,
  such as checkpoint/restore in userspace (CRIU),
  sleepable filter, and deployment configuration to make Seccomp-eBPF
  a privileged feature.
We also modified an existing container runtime (crun)
  to support Seccomp-eBPF-based system call filtering.

We use Seccomp-eBPF to implement various
  new security use cases, including
  system call count/rate limiting,
  flow-integrity protection (SFIP), and serialization.
We show how these features can prevent real-world vulnerabilities that cannot be
  safely prevented by cBPF filters.
We also use eBPF filters to enhance temporal specialization, which
  achieves up to 55.4\% reduction of the system call interface of the
  early execution phases,
  compared to existing cBPF implementations.
Lastly, we use eBPF filters to implement validation caching which
  can improve application performance by up to 10\%.

The paper makes the following main contributions:

\begin{itemize}
    \item We discuss several essential use cases of system call filtering,
      which reveal limitations of the state-of-the-art filtering mechanism exemplified
      by Seccomp.
    \item We present the design and implementation of Seccomp-eBPF
      that enhances programmability of system call security and integrates well with the
      kernel, without affecting performance or security.
    \item We implement and evaluate the advanced system call
      security features using Seccomp-eBPF filters
      for real-world applications against real-world vulnerabilities.
\end{itemize}

The code of our implementation of Seccomp-BPF on Linux 
  can be found at, 
\begin{center}  
\url{https://github.com/xlab-uiuc/seccomp-ebpf-upstream}
\end{center} 



\section{Background}
\label{sec:background}

In this section, we briefly discuss the necessary background
  for understanding the limitation of Seccomp
  as the state-of-the-practice
  system call filtering mechanism.






\subsection{Seccomp-cBPF}
\label{sec:background:cbpf}

Seccomp currently relies on the {\it classic} BPF (cBPF) language for users to express
  system call filters as programs~\cite{edge:lwn:2015,Kerrisk:2015}.
cBPF has a very simple register-based instruction set, making the filter programs easy to verify.
Due to the limited programmability,
  cBPF filters in Seccomp mostly implement an allow list---the filter only allows a system call if the system call ID is specified.  Occasionally,
a cBPF filter will further check arguments of primitive data types
  and prevent a system call if the argument check fails.
Pointer-typed arguments, however, cannot be dereferenced.

cBPF filters are {\it stateless}---the
  output of a Seccomp-cBPF filter execution depends only on the specified system call ID
  and argument values (in the allow list),
  because cBPF does not provide any utility of state management.

A cBPF filter is size-limited by 4096 instructions;
therefore, complex security policies have to
  be implemented by a chain of multiple filters.
All installed filters in the chain
  are executed for every system call, and the action with
  the highest precedence is returned.
The chaining behavior, however, comes with a performance overhead
  mainly due to security mitigations (e.g., Spectre)
  against indirect jumps~\cite{kuo:eurosys:22}.



Under the cover, Seccomp chooses to transform cBPF filters into eBPF
code internally to take the advantage of extensively optimized eBPF
toolchains, which produce code that runs up to 4X faster on x86-64
over using cBPF directly~\cite{Starovoitov:2014:patch}.
Note: this does not mean system call filters can be implemented in eBPF---the language
  exposed is still cBPF and eBPF features (e.g., maps and helpers) are not available.


\subsection{Seccomp Notifier}
\label{sec:background:notifier}

The limited expressiveness of cBPF makes it hard to implement complex security policies.
Therefore, Seccomp recently incorporated support for
  a userspace agent (called Notifier~\cite{utrap:brauner:blog})
  to complement cBPF filters.
Similar to early system call interposition frameworks~\cite{Garfinkel:2004,Garfinkel:2003,Wagner:1999,Goldberg:1996},
  it defers the decision to a trusted user agent.
Specifically, when Seccomp captures a system call, it blocks the calling task and
  redirects the system call context (e.g., calling PID, system call ID, and
  argument values) to the agent.


A major disadvantage of Seccomp Notifier
  is its significant performance overhead due
  to the additional context switches introduced by switching to
  the user space back and forth.
  Furthermore, to examine the contents of system call arguments that
  are user-space pointers, Seccomp Notifier must use \texttt{\small ptrace} to
  access the memory of the monitored process.
In addition to its
  performance implications, such inspection is subject to time-of-check-to-time-of-use (TOCTTOU) race
  conditions where a thread in the monitored program may change memory
  contents (and thus argument values) after the check has been
  completed by Seccomp Notifier.
Finally, the need to run a userspace agent in a trusted domain makes
  it challenging to be used in some deployment environments, e.g., for daemonless container runtimes~\cite{podmanio}.
  For all of these reasons, Seccomp Notifier is inadequate for complex system call filtering policies.


\section{Essential System Call Filtering Features}
\label{sec:usecases}

We highlight four features, none of which can be
implemented using Seccomp-cBPF, that will lead to a more effective, efficient,
and robust system call filtering framework:
statefulness, expressiveness, synchronization, and safe user memory access.
We motivate the
need for each feature with concrete examples.



\subsection{Statefulness}
\label{sec:usecases:state}
System call filtering with the existing Seccomp-cBPF framework is
fundamentally stateless: an invocation of the system call filter
cannot carry state to subsequent invocations.  As a result, policies
are overly loose. Here we describe three practical, tight policies
that require state passing: system call count limiting, system call
sequences checking, and enhanced temporal specialization.

\para{\bf System call count limiting.} Many attacks rely on
  invoking specific system calls repeatedly
  to cause starvation (e.g., \texttt{\small fork} bomb)
  or overflows (\S\ref{sec:eval:count}).
An effective defense is to limit the times a system call
  can be executed based on the demand of the application.

  As a concrete example, a special case of system call count limiting
  is to restrict the use of a specific system call to a single invocation,
  which is a common goal for container runtimes.
The goal is to prevent the launched containers from
  issuing \texttt{\small exec()} to replace the process image.
Container runtimes
take three common steps to launch a container:
1) installing Seccomp filters,
2) dropping privileges, and
3) launching containers using an \texttt{\small exec} system call.
The three steps are exemplified by the following code from runc~\cite{runc_code},
  a container runtime used by Docker and Kubernetes:
\begin{lstlisting}[language=myGo]
// Without NoNewPrivileges, seccomp is a
// privileged operation, so we need to do
// this before dropping capabilities
if l.cfg.Config.Seccomp != nil &&
    !l.cfg.NoNewPrivileges {
  seccomp.InitSeccomp(...)
}
// Drops the capabilities, sets the correct
// user and working dir, before executing the
// command inside the namespace
finalizeNamespace(...)
...
pdeath.Restore()
...
if unix.Getppid() != l.parentPid { ... }
...
unix.Write(fd, []byte("0"))
...
system.Exec(name, l.cfg.Args[0:], os.Env())
\end{lstlisting}


The snippet shows that the Seccomp filter needs to be installed \textit{before} calling \texttt{\small exec} (L19).
The desired policy is to only allow \texttt{\small exec} once at L19, but not later.
However, it is hard to implement the policy using stateless filters.
Note that the filter installation (L6) cannot be moved to a later point,
  because it requires the \texttt{\small CAP\_SYS\_ADMIN} capability, which is dropped within \texttt{\small finalizeNamespace} (L11).

As a result, with cBPF filters, the dangerous \texttt{\small exec} system call is commonly
  allowed for the entire duration of the container execution, even though it is not needed
  by the containerized applications~\cite{docker_seccomp_1,k8s_crun_seccomp}.

In fact, \texttt{\small exec} is not an exception.
The container runtime has to allow many other security-sensitive system calls in the same way,
  such as \texttt{\small prctl}, \texttt{\small capset}, and \texttt{\small write} as shown in the code snippet.
As with \texttt{\small exec}, these system calls cannot be blocked in a stateless filter, either.

If the system call filtering framework supported stateful policies,
the filter could keep track of the number of invocations of a
(potentially dangerous) system call and block further invocations.




\para{\bf Checking system call sequences and state machines.}
Recent work~\cite{Canella:2022} shows that an application's
  system call behavior can be
  modeled by a ``system call state machine'',
  which can be used for security
  enforcement named SFIP (Syscall Flow Integrity Protection);
  the state machine can be automatically generated using static analysis~\cite{Canella:2022}.
Moreover, prior work on IDS (Intrusion Detection Systems)
  has showed that
  an attack can be modeled by a sequence of system calls~\cite{Maggi:2010,Linn:2005,Warrender:1999,Forrest:1996,Feng:2003,Ghosh:1999}.
Support for stateful filters that maintain sequences and state machines would enable
a system call filtering framework to implement
both SFIP and IDS enforcement.

\para{\bf Precise temporal specialization.}
Prior work~\cite{Ghavamnia:usec:20,Lei:2017,Gu:dsn:14}
    shows the benefits of fine-grained security policies for
    different execution phases of the target application which often need distinct sets of system calls.

Ghavamnia et al.~\cite{Ghavamnia:usec:20} propose to apply cBPF Seccomp
  filters at different execution phases to achieve temporal system call specialization.
However, this technique is fundamentally limited under the current Seccomp-cBPF
  security model.
In Seccomp, a filter, once installed, cannot be uninstalled during the process lifetime.
Filters installed in later phases are chained with the filters installed earlier;
all the installed filters are executed for every system call and the most restrictive policy is applied.
Hence, with cBPF Seccomp filters,
  a system call needed in Phase $N$ has to be allowed in all earlier phases (Phases $1...N-1$),
  even though the system call is not needed in any of the $N-1$ phases.
Figure~\ref{fig:tss} illustrates this point with a two-phase temporal specialization ($N=2$),
  where P1 (Initialization) has to include system calls from P2 (Serving) with Seccomp-cBPF.

The limitation is rooted in the fact that cBPF filters cannot record states
  (i.e., the current execution phase).
By recording the current phase in a state variable and applying the corresponding policy,
  a stateful system call filtering framework would enable
  precise temporal specialization and achieve tighter security policies for each phase.



\subsection{Expressiveness}
\label{sec:usecases:expressive}
As with any user-supplied code or policy running in the kernel, system
call filtering frameworks typically must trade off expressiveness and
safety.  The safety goals of cBPF have led it to a design prioritizing
safety, with an overly restricted instruction set and runtime.

\para{Performance optimizations.}
Given the cBPF instruction set,
    a cBPF filter is typically in the form of an allow list,
    implemented by a series of conditional jumps~\cite{edge:lwn:2015,Kerrisk:2015}.
Complex policies 
  result in long lists of jumps and multiple filters (due to the size
  limit of a cBPF filter, see \S\ref{sec:background}).
Consequently, system call checking becomes expensive due to the
  need of iterating over long jump lists~\cite{Hromatka:2018,Hromatka:2018:2,skarlatos:micro:20,ncc:2016,Kerrisk:2015}
  and the overhead of indirect jumps caused by mitigations
  to speculative vulnerabilities (e.g., Retpoline)~\cite{kuo:eurosys:22}.
To reduce the overhead, multiple optimizations were proposed, such as
    dedicated Seccomp caches~\cite{skarlatos:micro:20}, skip-list search~\cite{DeMarinis:20},
    and filter merging~\cite{kuo:eurosys:22}.

A more expressive system call filtering framework
  can optimize the filter performance.
First, advanced data structures with constant lookup time (e.g., a hash map)
  can be used to eliminate long jump lists.
Such an implementation is essentially equivalent to the dedicated
    Seccomp cache~\cite{skarlatos:micro:20}.

Moreover, cBPF filters are limited by 4096 instructions; allowing more instructions
could eliminate the overhead of indirect jumps caused by chaining
  multiple filters, raised by Retpoline
  or other mitigations to speculative vulnerabilities.
In a stateful system call filtering environment,
  one can further use map entries to sequence filters for more complex policies \newjinghao{with low overhead}.
As we will see in \S\ref{sec:design}, such enhancements to expressiveness can be achieved
without sacrificing safety.

\para{Rate limiting.} \new{Besides the performance aspects,
  the limited expressiveness of cBPF,
    due to the simple instruction sets
    and the program size constraints,
    also makes it hard to support advanced policies.
One such example is rate limiting which only allows
  specified system calls to be issued under expected rates.
Rate limiting relies on a timer. However,
  there is no timer utility in cBPF; in fact, cBPF
  cannot invoke any kernel functions or utilities.
Furthermore, cBPF cannot record the last time
  in any state variable.
A system call filtering framework with better expressiveness could be a
  solution; in particular, by exposing advanced and complex operations to the
  filters in a safe way. Such a framework could expose the current time
  information to the filter to facilitate the desired system call rate limiting.}



\subsection{Synchronization}
\label{sec:usecase:serialization}

\new{System call filtering frameworks essentially provide a platform for
  mandating access into the kernel and can be used to implement application- or
  system-wide policies to prevent misuse of the kernel.}
\new{Specifically, there has been increasing reports
  on kernel vulnerabilities that manifest via race conditions
  and are exploitable by
  two concurrently executing system calls~\cite{Zhao:dsn:18,Jeong2019razzer,Watson:2007,Gong:sosp:21,Xu:2020}.
Mitigating such attacks requires kernel developers
  to identify the race condition in the kernel,
  patch the vulnerable code,
  potentially backport the revisions to older kernel versions,
  and release a patched version.}
The above process is time-consuming; waiting for
  the kernel patches could open a long window of vulnerability.
A system call filtering framework which can serialize
  specific system calls that are known to be exploitable
    by system call racing
  can immediately and effectively nullify
  race conditions without waiting for patches, backports, and releases.

\subsection{Safe User Memory Access}
\label{sec:usecases:dpi}

\new{Since a large number of system calls take pointers to user memory as arguments,
  deep argument inspection (DPI) is long desired~\cite{Edge:deeparg:lwn,edge:lwn:2020}.
Seccomp-cBPF cannot support DPI because it only checks
  non-pointer argument values, i.e.,
  if an argument is a pointer, it cannot be dereferenced by Seccomp-cBPF,
  which means that accepting or rejecting the system call cannot depend on
  values in structures that are passed to system calls via pointers.
In fact, it means that Seccomp-cBPF cannot even address any string values
  (e.g., a file path in \texttt{\small open()}).
To enable DPI,
  a system call filtering framework needs to provide
  a safe way to access user memory referred to by the pointer arguments.
The main challenge (which is also the reason that Seccomp does not dereference pointers)
  is to avoid the time-of-check-to-time-of-use (TOCTTOU)
  issue~\cite{Garfinkel:2003,Bhattacharyya:usec:22,Payer:vee:12}, where user space can change the value of what is being pointed to
  between the time the filter checks it and the time the value gets used.}


\section{Threat Model and Design Goals}
\label{sec:overview}


\subsection{Threat Model}
\label{sec:threat_model}

We \new{strictly adhere to} the current threat model of Seccomp.
The goal is to restrict how untrusted userspace applications interact with the shared OS kernel
    through system calls to protect the kernel from userspace exploits (e.g., shellcode or ROP payload).
The kernel is trusted.


Seccomp requires the calling context to either be privileged (having \texttt{\small CAP\_SYS\_ADMIN}
    in its user namespace),
    or set \texttt{\small NO\_NEW\_PRIVS}~\cite{Edge:set_no_new_priv:lwn} which
    ensures that an unprivileged process
    cannot apply a malicious filter
    and then invoke a set-user-ID or other privileged program using \texttt{\small exec}.


%
Once a filter is installed onto a process,
    it cannot be removed before the process termination.
A filter cannot be tampered---a filter program and
    its states
    will be invisible to unprivileged processes once it is installed. 

\comments{
Same as cBPF filters, once an eBPF filter is installed onto a process,
    it cannot be removed before the process termination.
An eBPF filter cannot be tampered with by other applications---a filter program and
    its states in maps
    will be invisible to unprivileged processes, once installed. 
}

\comments{
The support of eBPF Seccomp filters can be configured to be a privileged feature (root only),
    or an unprivileged feature.
For the unprivileged configuration, unprivileged user processes can potentially
    load and install a malicious filter to exploit vulnerabilities in the Seccomp or
    eBPF subsystems.
Note that the same attack path might also exist with cBPF filters (as cBPF code
  is transformed into eBPF, see \S\ref{sec:background:cbpf}).
}



\subsection{Design Goals}
\label{sec:high-level-overview}

Given this threat model, we set the following goals:
\begin{itemize}
\item{\bf Expressiveness and kernel support.}
\new{We aim to support all the essential system call filtering
  features discussed in \S\ref{sec:usecases}.
This not only requires a more programmable and expressive language,
  but also additional kernel support.}
\item{\bf Maintain Seccomp usage model and interfaces.}
  In order to provide a practical, familiar and useful system call filtering framework,
  we must adhere to the same
  usage and threat model of Seccomp (\S\ref{sec:threat_model}).
  Further, to lower the barrier of adoption, we aim to maintain its interface.
\item{\bf No privilege escalation for unprivileged users.}
  As Seccomp supports the unprivileged use case (\S\ref{sec:threat_model}),
  our design must ensure no privilege escalation.
\end{itemize}

\comments{
\para{Security as a first-class principle.}
Supporting the new Seccomp eBPF program type should not reduce the security of the system.
Our design and implementation adhere to the security models of
  the two subsystems that Seccomp-eBPF builds upon: Seccomp and eBPF.
Seccomp-eBPF does not allow a potential
  attacker to gain information
  that would not be accessible with other eBPF programs (e.g., socket filters).
\comments{
Seccomp requires \texttt{\small CAP\_SYS\_ADMIN} or \texttt{\small NO\_NEW\_PRIVS},
which sets the context for eBPF filters (see \S\ref{sec:threat_model}).

We maintain the capability requirements of the existing eBPF
helper functions when exposing them to eBPF filters.

For our newly created helpers,
  we strictly define the permissions to prevent exposure of critical information.

We implement a \texttt{\small sysctl} configuration for setting eBPF Seccomp filter as a privileged feature in case that unprivileged eBPF is a concern.
}

\para{Not breaking Seccomp-cBPF.}
The new program type should co-exist with Seccomp-cBPF.
Applications that do not require advanced filters can continue using cBPF filters.

\comments{
cBPF filters fit simple policies that can be expressed using static allow/deny lists.

Nevertheless, the advantages of eBPF filters over cBPF filters are fundamental.
Not only does it provide an extended language, but also the ability to store states in maps,
  and to invoke and extend in-kernel helper functions.
Hence, Seccomp-eBPF provides new mechanisms to enable more advanced security features (\S\ref{sec:usecases}).
}


\comments{
The challenge lies in the different load and install procedures
  for cBPF filters and eBPF filters.
Our implementation maintains the \texttt{\small seccomp()} interface as far as possible, including both its signatures and the parameter semantics.
}
}

\section{Design}
\label{sec:design}


\subsection{Overview}
\label{sec:design:basic}

\new{We develop a programmable system call filtering mechanism
  on top of Seccomp,
  by leveraging the {\it extended} BPF language (eBPF),
  to enable advanced security policies (see \S\ref{sec:usecases}).
eBPF is a fundamental redesign of the BPF infrastructure within the Linux kernel~\cite{ebpfio}.
It not only has a rich instruction set and flexible control flows (e.g., bounded loops
  and BPF-to-BPF calls), but also
    offers new features, such as {\it helper functions}
    to interface with kernel utilities and
    {\it maps} as efficient storage primitives to maintain states.}




\new{The choice of eBPF is a result of practicality considerations:
  1) eBPF offers basic building blocks
    for the target programmability, including maps for statefulness
    and helper functions for interfacing with the kernel;
  2) eBPF is verified to be safe by the kernel;
  and 3) since Seccomp already converts cBPF code into eBPF
    internally (\S\ref{sec:background:cbpf}),
    we can largely reuse existing Seccomp implementation and workflow.}

\new{However, directly opening the eBPF interface in Seccomp is not a solution:
  1) basic supports are missing (e.g., synchronization primitives for serialization);
  2) existing utilities (e.g., task storage) may not fit the Seccomp model because
    they target privileged context only;
and 3) existing features are not safe for Seccomp use cases, e.g., the
  current user memory access feature cannot address TOCTTOU issues.}

\begin{figure}
  \centering
  \includegraphics[width=0.5\textwidth]{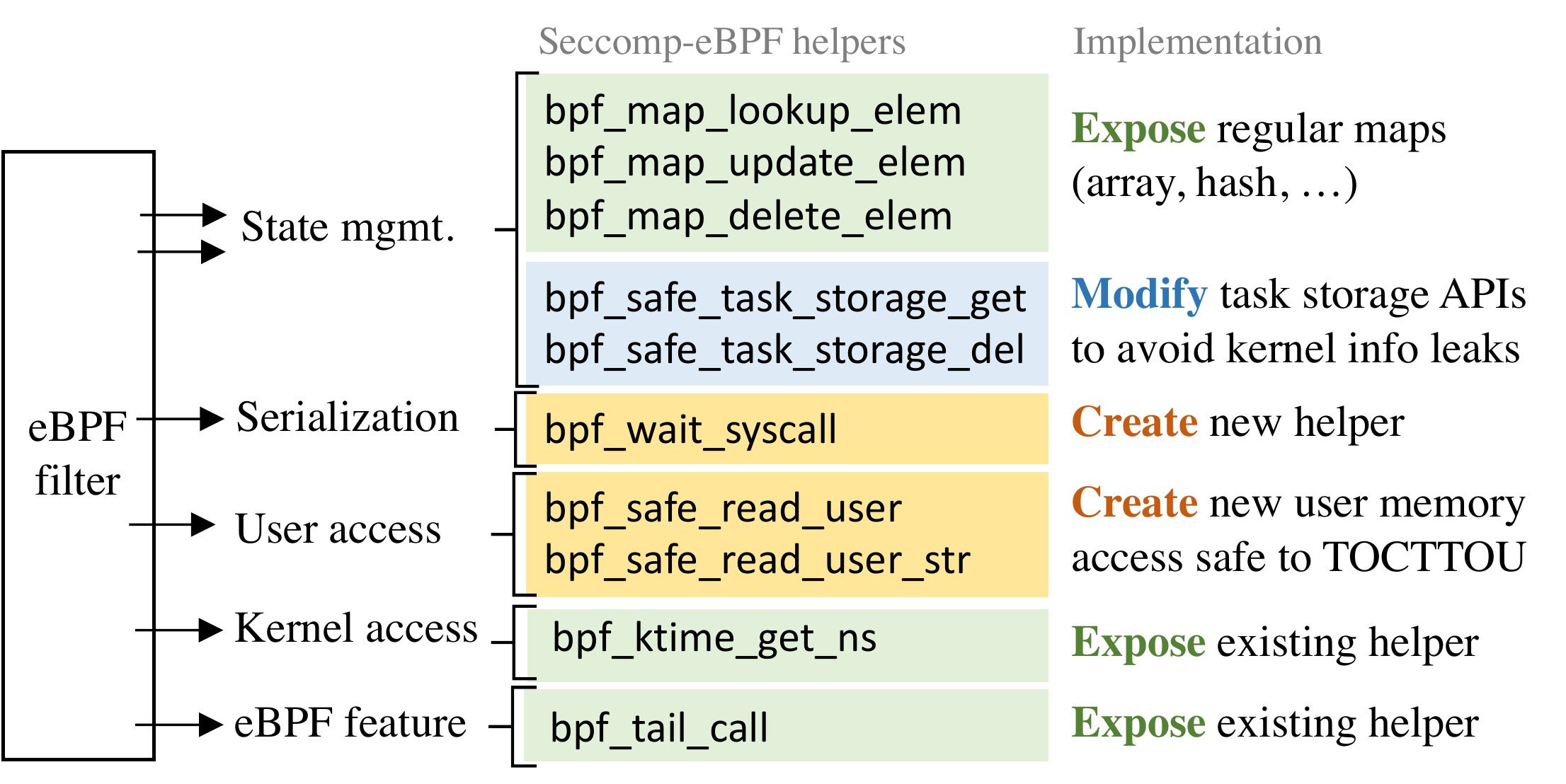}
  \vspace{-15pt}
  \caption{Features, helper interfaces, and their implementation of
    the Seccomp eBPF program type.}
  \label{fig:helpers}
\end{figure}

\new{We expose, modify, and create new eBPF helper functions to
  safely manage filter state, access kernel and user state, and
  utilize synchronization primitives.
Figure~\ref{fig:helpers} illustrates these helper functions.
Importantly, our system
  integrates with existing kernel privilege and capability
  mechanisms, enabling unprivileged users to safely install advanced filters.
Essentially, we create a new {\em Seccomp-eBPF} program
  type which is highly programmable to
  express advanced system call security policies
  in eBPF filter programs.}

\new{In terms of security, we reduce the security of Seccomp-eBPF to the security
  of Seccomp and eBPF.}


\subsection{Seccomp-eBPF Program Type}

\begin{figure}
  \centering
  \includegraphics[width=0.48\textwidth]{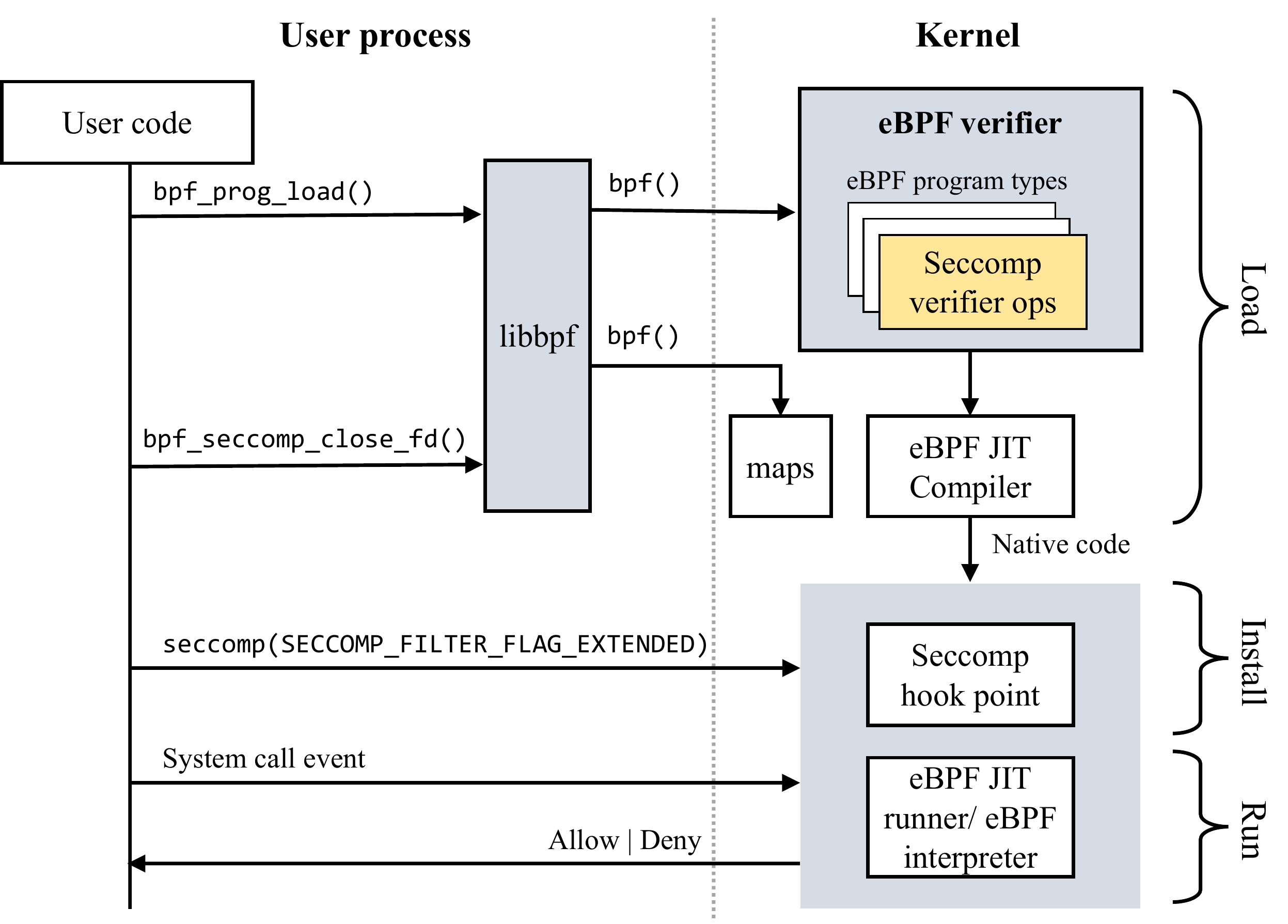}
  \caption{Workflow of a Seccomp-eBPF filter}
  \vspace{-5pt}
  \label{fig:workflow}
\end{figure}

\new{An eBPF program type defines what helper functions (helpers) a program of the type
  is allowed to invoke and the corresponding capabilities
  required for invoking them.
In other words, a program type defines its interface to the kernel
  and the capability system to use the interface.
For a given filter program,
  the eBPF verifier checks the helper invocation instructions in the
  filter program and the
  capabilities of the calling context at the load time,
  as shown in \cref{fig:workflow}.}




\new{Specifically, the eBPF verifier checks
  both eBPF language specifications
  and  Seccomp-eBPF program type.
We extend the eBPF verifier to check the new Seccomp-eBPF
  program type, \texttt{\small BPF\_PROG\_TYPE\_SECCOMP}.
Seccomp-eBPF restricts
  (1) the use of helper functions,
  (2) the access to Seccomp data structures.}
\newjinghao{The eBPF verifier already provides hooks where we
  directly add our verification code.
For (1), we declare the helper functions that eBPF filters are permitted to
  use (\S\ref{sec:design:helper}).}
For (2), we verify that filters only access data within the boundary of Seccomp data
  structures with correct offset and size.
We strictly follow the verification against eBPF language specifications.

\new{\cref{fig:workflow} depicts the workflow of loading, installing, and running
  a Seccomp-eBPF filter.
The filter is first loaded into the kernel, where it is verified
    and optionally JIT-compiled,
  and then installed in Seccomp.
After that, the installed eBPF filter
  will be invoked upon every subsequent system call.}






\comments{
The ``base'' helpers (for map access, tail calls, and timer) can be invoked by unprivileged eBPF programs
  (eBPF code loaded from an unprivileged context).
Helpers for spin locks and CPU structures require additional capability (\texttt{\small CAP\_BPF});
  and hence can only be used by privileged eBPF.
The other helpers additionally require \texttt{\small CAP\_PERFMON}.
We adhere to these capability requirements when exposing those helpers to Seccomp-eBPF filters.
}



\comments{
\begin{table*}
   \footnotesize
   \centering
   \begin{tabular}{lll}
     \toprule
     & {\bf Helper Function}\\ 
     \midrule
      Existing & \texttt{void *bpf\_map\_lookup\_elem(struct bpf\_map *map, const void *key)}\\ 
      Existing & \texttt{long bpf\_map\_update\_elem(struct bpf\_map *map, const void *key, const void *value, u64 flags)}\\ 
      Existing & \texttt{long bpf\_map\_delete\_elem(struct bpf\_map *map, const void *key)}\\ 
      Existing & \texttt{long bpf\_tail\_call(void *ctx, struct bpf\_map *prog\_array\_map, u32 index)}\\        
      Existing & \texttt{u64 \  bpf\_ktime\_get\_ns(void)}\\    
      New & \texttt{long bpf\_safe\_read\_user(void *dst, u32 size, const void *unsafe\_ptr)}\\        
      New & \texttt{long bpf\_safe\_read\_user\_str(void *dst, u32 size, const void *unsafe\_ptr)}\\   
      New & \texttt{void *bpf\_safe\_task\_storage\_get(struct bpf\_map *map, struct task\_struct *task, void *value, u64 flags)}\\    
      New & \texttt{long bpf\_safe\_task\_storage\_delete(struct bpf\_map *map, struct task\_struct *task)}\\ 
      New & \texttt{void bpf\_wait\_syscall(int this\_nr, int target\_nr)}\\          
     \bottomrule
   \end{tabular}
   \caption{Helper functions that are provided for Seccomp-eBPF filters.\tianyin{really need to finalize the table,
    can perhaps change them to the full signature}}
   \label{table:base_func_helper}
 \end{table*}
}

\subsection{Helper Functions}
\label{sec:design:helper}

\new{Figure~\ref{fig:helpers} shows the helper functions of the Seccomp-eBPF
  program type (\texttt{\small BPF\_PROG\_TYPE\_SECCOMP}).
Overall, there are ten helpers in five categories:
  (1) {\it state management} based on helpers to access (lookup/update/delete) maps,
  (2) {\it serialization} which needs synchronization helpers,
  (3) {\it user access} for safely accessing user memory,
  (4) {\it kernel access} for invoking kernel utilities such as time,
  and (5) {\it eBPF program features} such as tail calls that allow more complex programs.}

\new{Five of them already exists in Linux and can be directly {\it exposed}
  to the Seccomp-eBPF program type.
The others are not available. In this section,
  we discuss the {\it new} helpers
  that are created from scratch or by modifying existing ones.}



\subsubsection{Task Storage}
\label{sec:design:task_storage}
Maps are the basic enabler to statefulness (\S\ref{sec:usecases:state}).
eBPF currently supports different types of maps,
  including array maps and hash maps.

One important map type is task storage, which provides
  primitives for policies
  that need to maintain separate states for different
  tasks executing the same eBPF filter (loaded by the same FD).
eBPF already implements task storage map, which
  uses Linux tasks as keys to access the stored values.
This guarantees that storage from different tasks does not collide.
The task storage can be used through two helpers:
(1) \texttt{\small bpf\_task\_storage\_get} 
and (2) \texttt{\small bpf\_task\_storage\_delete}. 

However, both the two helpers
  require additional capabilities, \texttt{\small CAP\_BPF} and \texttt{\small CAP\_PERFMON},
  and thus cannot be used in unprivileged contexts,
  which hinders Seccomp's use cases
  (an unprivileged user process can install
  a Seccomp filter as long as it sets \texttt{\small NO\_NEW\_PRIVS},
  see \S\ref{sec:threat_model}).
\new{
Our goal is to redesign the helpers for task storage maps
  to make them essential utilities of Seccomp-eBPF filters.
}



\new{Why do the helpers for task storage maps need additional privileges, while the
  other map helpers
  do not (e.g., \texttt{\small bpf\_map\_lookup\_elem} in Figure~\ref{fig:helpers})?
The reason is that the API design
  of the existing helpers for task storage maps
  makes them insecure to unprivileged Seccomp-eBPF filters:}
  the existing helpers used to access the
  task storage require a \texttt{\small task\_struct} pointer as input argument to
  find the corresponding map storage.
While a privileged eBPF filter can call
  \texttt{\small bpf\_get\_current\_task\_btf} to retrieve the needed \texttt{\small task\_struct}
  pointer, such operation is insecure in an unprivileged context because
  a \texttt{\small task\_struct} contains a pointer to its parent
  \texttt{\small task\_struct}.
A malicious filter can dereference the parent pointer
  recursively to reach the init task (PID 0) from the current process,
  leaking sensitive kernel information.



To avoid this issue and provide
unprivileged filters with secure task storage, we create a
new set of task-storage helpers.
Our new helpers do not require a
  \texttt{\small task\_struct} as input.
Instead, the helpers automatically find the current task's
  group leader.
For a process, this is its own \texttt{\small task\_struct}; for a thread, this is the
  group leader---the task that spawned it.

\subsubsection{Safe User Memory Access}
\label{sec:design:user_memory_access}

Safely accessing user memory is an important feature for checking pointer-typed argument values,
  as discussed in \S\ref{sec:usecases:dpi}.
We support such a feature and expose it through specialized helpers.
Note that eBPF provides two helpers for accessing memory of user processes.
However, these helpers cannot be used for deep argument inspection because
 they cannot address TOCTTOU-based argument racing~\cite{Garfinkel:2003,Watson:2007}.

We follow the prior work on deep argument inspection~\cite{Canella_submission,Edge:deeparg:lwn}.
The key principle is to disallow user space to
    modify the argument values during and after the values are checked.
This can be achieved by
(1) copying the target user memory into a protected memory region that is only accessible by the kernel,
(2) making the target user memory write-protected or inaccessible to user space, or
(3) using protection-key functionalities in the kernel to prevent races from user space~\cite{Corbet:key:2020,Gravani:21,Frassetto:2018}.
We implement the first two solutions, as they can be done on all commodity
  hardware, and expose them with user-memory access helpers.

\para{Capability.} The existing user-memory access helpers from eBPF
  require \texttt{\small CAP\_BPF} and \texttt{\small CAP\_PERFMON}.
We reduce the security of Seccomp-eBPF to Seccomp Notifier~\cite{utrap:brauner:blog} which allows
 the userspace agent to read and copy user memory if
 the agent is allowed to \texttt{\small ptrace} the process.
  i.e.,
  the security of Seccomp Notifier is equal to \texttt{\small ptrace}.
Therefore, we also reduce the capabilities of user-memory access helpers
  to \texttt{\small ptrace}.

Linux determines if one process (tracer) can \texttt{\small ptrace} another
  process (tracee) based on the following checks:
1) they are in the same thread group,
2) they are under the same user and group, or if the tracer has \texttt{\small CAP\_SYS\_PTRACE},
3) the tracee cannot be traced without \texttt{\small CAP\_SYS\_PTRACE} if it has set itself to be non-dumpable, and
4) other LSM hooks.


How does \texttt{\small ptrace} apply to Seccomp?
In Seccomp, the filter is
  regarded as the tracer, with the process that installs the filter
  being the tracee.
Since unprivileged Seccomp requires the \texttt{\small NO\_NEW\_PRIVS} attribute
  to be set on the calling task, the UID/GID and capability set can not
  change after the filter installation.
Hence, the in-kernel \texttt{\small ptrace} checks are largely
  covered by the \texttt{\small NO\_NEW\_PRIVS} attribute.
Furthermore, the dumpable attribute is respected.

On Linux, the capability of \texttt{\small ptrace} also depends on a system-wide
  configuration, \texttt{\small ptrace\_scope}~\cite{ptrace:scope:yama}.
To ensure that user-memory access helpers follow \texttt{\small ptrace} security,
  we define a new LSM hook that enforces the helpers to adhere to the
    policy set by \texttt{\small ptrace\_scope}.


\para{Protecting non-dumpable process.}
On Linux, a process (e.g., OpenSSH) handling sensitive information can set the ``dumpable'' attribute
  (via \texttt{\small PR\_SET\_DUMPABLE})
  to prevent being coredumped or ptraced
  by another unprivileged process. 
Hence, an unprivileged Seccomp-eBPF filter should not be able to
    access memory of non-dumpable processes.
To protect such processes, we apply a privilege requirement similar to ptrace.
The helpers are modified to allow a filter to access non-dumpable memory only if the loader
  process has ptrace privileges (\texttt{\small CAP\_SYS\_PTRACE}).


\para{Handling page faults.}
The original user-memory access helpers in Linux do not
  handle page faults
  when reading memory from userspace.
The helpers use non-blocking functions
  (e.g., \texttt{\small copy\_from\_user\_nofault}),
  which emit errors upon page faults.
This is based on the dated assumption that a BPF program never sleeps,
  i.e., it cannot be blocked in the middle of execution~\cite{sleepable:bpf:lwn}.
This is inconvenient for Seccomp-eBPF filters---invoking a user-memory helper would fail if the memory access triggers a page fault.
We support {\it sleepable} Seccomp-eBPF filters (\S\ref{sec:impl}).
Therefore, our user-memory access helpers can handle page faults.

\subsubsection{Serializing System Calls}
\label{sec:syscall_serialize}

\comments{
  To demonstrate this point,
  we enable eBPF filters to implement global system call serialization (\S\ref{sec:usecases})
  by creating a new helper, using which one would be able to serialize certain
  system calls and mititgate vulnerabilities that manifests from race conditions
  between system calls on this system.}

\new{The basic idea to serialize two racing system calls
  is to record the event of involved system calls.
When a system call is invoked concurrently with another system call under
  execution which is known to have race vulnerabilities with it,
  the kernel stops the former system call until the latter finishes.}

\comments{ 
\new{System call serialization can be supported per-process or
  system wide. For per-process system call serialization,
  we extend \texttt{\small task\_struct} with a new struct-typed field to
  store the ID and arguments of target system calls that
  are executing on the task.
For system-wide serialization, we \tianyin{XXX @Jinghao: write about how
  do you record the execution.}}
}

\new{In this use case, the kernel records whether a particular system call (in terms
  of system call ID) is currently being executed and stores the information.
For each system call, we add an integer atomic variable in the kernel to store
  whether there exists processes that are currently executing the system call.
This atomic variable has an initial value of 0 and will be incremented by
  our new helper function.
When a system call exits, its atomic variable is decremented, but the value will
  have a minimum of 0.}


\para{Helper API.}
\new{We expose a new helper function,}

  \texttt{\small void bpf\_wait\_syscall(int curr\_nr, int target\_nr)}

\noindent
\new{for system call serialization.
The helper function holds the execution of the current task until
  the execution of the target system call finishes.
To achieve this, it increments the atomic variable of the current system call
  and perform busy waiting via a schedule loop until the atomic variable of the
  other system call is decremented to 0, which means no processes are
  executing the other system call.
}
\new{The helper is unprivileged, because it only affects the
  processes that attach the filter.}

\para{Enforcement.}
\new{The eBPF filter that implements serialization uses a map to store system calls
  that have race vulnerabilities as key-value pairs.
The map can be updated by a trusted userspace process after filter
  installation, when new race vulnerabilities
  need to be patched (\S\ref{sec:usecase:serialization}).
The filter implements a logic that, for each incoming system call, it
  checks whether the system call has a potential race condition based on the map.
If so, the filter will invoke the \texttt{\small bpf\_wait\_syscall}
  helper to serialize the current system call.}

\new{To enforce system-wide serialization policies,
  we install the filter onto
  the \texttt{\small init} process.
In Seccomp, a process inherits the filter of its parent process.
Since \texttt{\small init} is
  the ancestor of all subsequent processes,
  the installed filter is inherited by all processes on the system,
  hence implementing a global policy.
Since the kernel requires the root privilege when retrieving map file descriptor,
  only trusted, privileged processes can update the map.}

\subsection{Usage}
\label{sec:usage}
The user loads a Seccomp-eBPF filter into the kernel via the \texttt{\small bpf()} system call
  and installs it using the \texttt{\small seccomp()} system call.
Different from cBPF filters which are implemented using BPF instructions,
  the kernel expects eBPF filters to be loaded as bytecode.
This allows eBPF filters to be written in high-level languages, such as C and Rust,
  and compiled using LLVM/Clang.
In fact, eBPF has more mature and actively developing tool chains than cBPF
  (cBPF ``{\it is frozen}~\cite{Starovoitov:2020:seccomp_future_dev}.'').

We add a new flag,
  \texttt{\small SECCOMP\_FILTER\_FLAG\_EXTENDED},
  to the \texttt{\small seccomp()} system call;
  if it is set,
  the filter is interpreted in eBPF; otherwise, it is in cBPF.
Note that, different from cBPF filters,
  before installing an eBPF filter,
  one needs to load it using \texttt{\small bpf()}.
Figure~\ref{fig:ld} shows the code snippets of loading and installing
  a Seccomp-eBPF filter, compared with installing a Seccomp-cBPF filter.

\begin{figure}
  \centering
  \begin{subfigure}[b]{0.5\textwidth}
\begin{lstlisting}[language=myJava,label={lst:interface}]
struct sock_fprog prog = ...;
prctl(PR_SET_NO_NEW_PRIVS, 1, 0, 0, 0);
// Load and install the filter program in Seccomp
seccomp(SECCOMP_SET_MODE_FILTER, 0, &prog);
\end{lstlisting}
    \caption{cBPF: load and install in one step}
  \end{subfigure}
  \begin{subfigure}[b]{0.5\textwidth}
\begin{lstlisting}[language=myJava,label={lst:interface}]
// Load the eBPF filter in bytecode
bpf_prog_load(filter_path,
    BPF_PROG_TYPE_SECCOMP, &obj, &fd);
prctl(PR_SET_NO_NEW_PRIVS, 1, 0, 0, 0);
bpf_seccomp_close_fd(obj);
// Install eBPF filter in Seccomp
seccomp(SECCOMP_SET_MODE_FILTER,
    SECCOMP_FILTER_FLAG_EXTENDED, &fd);
\end{lstlisting}
    \caption{eBPF: load and install in two steps; verification invoked on load}
  \end{subfigure}
  \caption{Code snippet for installing Seccomp filters in (a) cBPF and (b) eBPF.
  \texttt{\small bpf\_prog\_load} is the libbpf function that wraps the \texttt{\small bpf()} system call;
  \texttt{\small bpf\_seccomp\_close\_fd} closes all file descriptors (FDs) except the eBPF program FD
  (which is needed as a parameter for \texttt{\small seccomp()} and gets closed inside).}
  \label{fig:ld}
\end{figure}

\new{Note that cBPF does not have a separate load step (despite its name, the
  \texttt{\small bpf()} system call is specific to eBPF).
  A cBPF filter is
  installed in one step through the \texttt{\small seccomp()} system call,
  where the verification is done at the installation time.
We choose to minimize changes to the existing \texttt{\small seccomp()} and \texttt{\small bpf()} interfaces,
  mainly for reducing adoption obstacles.
However, the separation requires additional efforts
  to protect filters and maps against tampering (\S\ref{sec:design:protection}).}

\subsection{Tamper Protection}
\label{sec:design:protection}

As a Seccomp-eBPF filter is verified and installed in two steps (\S\ref{sec:usage}),
 we develop the following two tamper protections.

\para{Protecting filter program and maps.}
For eBPF filters, \texttt{\small seccomp()} automatically closes the file descriptor (FD)
  of the eBPF program before returning to the user space.
The user space is responsible for closing all FDs of maps before issuing the \texttt{\small seccomp} system call.
This is to prevent leaks of the FDs, as anyone with access to the maps can potentially manipulate filter's behavior.
Our new function in libbpf, \texttt{\small bpf\_seccomp\_close\_fd}, closes all these FDs except the eBPF program FD,
  which is closed by \texttt{\small seccomp} itself.
The maps themselves are ref-counted by the eBPF filter program after eBPF
  verification; therefore, closing the map FDs does not remove the maps~\cite{Starovoitov_bpf_obj}.

\para{Namespace tracking.}
\new{Currently, none of the helper functions we expose to Seccomp-eBPF filters
  requires additional privileges.
If there is a need to expose helpers that require \texttt{\small CAP\_BPF} and/or
  \texttt{\small CAP\_PERFMON} (like many existing helpers),
  we need to enforce that
  an eBPF filter is loaded and installed in the same user namespace.
Otherwise,
    an unprivileged attacker 
    in the current user namespace can
    create a new user namespace (which has all capabilities by default~\cite{namespace:lwn})
    to bypass the capability checks in the verifier, and then sends the filter
    FD back to its restricted parent namespace).}
We develop the enforcement by recording a reference to the load-time user namespace with the filter.
When installing a filter, the kernel checks whether the current user
    namespace matches the recorded load-time user namespace.

\comments{
The namespace tracking can prevent potential verification bypasses (e.g.,
  an unprivileged attacker 
  in the current user namespace
  could create a new user namespace (which has all capabilities by default~\cite{namespace:lwn})
  to bypass the capability checks in the verifier, and then sends the filter
  FD back to its restricted parent namespace).
}

\comments{
\subsection{Security Analysis}
\label{sec:design:security}


\begin{itemize}
\item{\bf Seccomp.} \newjinghao{Our design keeps the same filter installation
  requirements of the existing Seccomp, where only processes with
  \texttt{\small CAP\_SYS\_ADMIN} capability or \texttt{\small NO\_NEW\_PRIVS}
  attributes are allowed to install a filter.}
All file descriptors used by an eBPF
  Seccomp filter are closed after the filter is installed onto the process.
This ensures that filters and maps cannot be leaked.
Moreover, we enforce eBPF filters to be loaded and installed in the
  same user namespace to prevent a process from installing
  a filter verified in a different namespace.

\item{\bf eBPF.} We do not relax any safety constraints enforced by the eBPF verifier---an
  Seccomp-eBPF filter is verified in the same way as any other eBPF program.
We do not modify the capabilities required
  by existing helpers.
\end{itemize}

\para{Task storage.}
Our new helpers for task storage maps (\S\ref{sec:design:map}) are considered to be secure.
  For task storage, we introduce two new helpers to allow unprivileged eBPF
  filters to maintain per-process states,
  without exposing sensitive information about the kernel or other processes.



\para{User-memory access.}
From the eBPF perspective, reading user memory is a change of security,
  because, currently, unprivileged eBPF programs cannot access user memory (it requires \texttt{\small CAP\_BPF} and \texttt{\small CAP\_PERFMON}).
However, reading and inspecting user-memory can be done by Seccomp Notifier~\cite{utrap:brauner:blog}.
Hence, we allow unprivileged Seccomp-eBPF filters to be able to access and inspect user memory.
We therefore reduce the security to Seccomp Notifier.

Seccomp Notifier adheres to strict security policies in terms of reading user memory,
  where the trusted userspace agent can inspect the target
   process memory
  to read the values referred to by pointer arguments.
To do so, the agent needs to have \texttt{\small ptrace} capabilities.
Therefore, the security of our user-memory access helpers is equal to \texttt{\small ptrace}.


Linux determines if one process (tracer) can \texttt{\small ptrace} another
  process (tracee) based on the following checks:
1) they are in the same thread group,
2) they are under the same user and group, or if the tracer has \texttt{\small CAP\_SYS\_PTRACE},
3) the tracee cannot be traced without \texttt{\small CAP\_SYS\_PTRACE} if it has set itself to be non-dumpable, and
4) other LSM hooks.


How does \texttt{\small ptrace} apply to Seccomp?
In Seccomp, the filter is
  regarded as the tracer, with the process that installs the filter
  being the tracee.
Since unprivileged Seccomp requires the \texttt{\small NO\_NEW\_PRIVS} attribute
  to be set on the calling task, the UID/GID and capability set can not
  change after the filter installation.
Hence, the in-kernel \texttt{\small ptrace} checks are largely
  covered by the \texttt{\small NO\_NEW\_PRIVS} attribute.
Furthermore, the dumpable attribute is respected (\S\ref{sec:design:user_memory_access}).

On Linux, the capability of \texttt{\small ptrace} also depends on a system-wide
  configuration, \texttt{\small ptrace\_scope}~\cite{ptrace:scope:yama}.
To ensure that user-memory access helpers follow \texttt{\small ptrace} security,
  we define a new LSM hook that enforces the helpers to adhere to the
    policy set by \texttt{\small ptrace\_scope}.


%
}

\section{Implementation}
\label{sec:impl}

For practical uses in a variety of deployment environments (such as 
    container environments), 
    we addressed a number of implementation challenges.

\para{Sleepable Seccomp filters.} 
Sleepable filters are needed for Seccomp-eBPF
  (cBPF programs do not need to sleep),
  e.g., for page fault handling when accessing user memory (\S\ref{sec:design:user_memory_access}).
To support sleepable filters, we create new BPF section names (\texttt{\small seccomp} and \texttt{\small seccomp-sleepable}).
The BPF section is an ELF section that is used by libbpf
  to determine the sleepable attribute of the BPF program and set the flag
  when calling \texttt{\small bpf()}.
This allows us to maintain the same \texttt{\small bpf()} interface without additional flags or other tooling support.

\para{Checkpoint/restore in userspace (CRIU).}
CRIU is widely used by container engines to checkpoint
  the state of a running container to disk and restore it later.
It facilitates features such as live migrations or snapshots.
Seccomp currently supports CRIU only for cBPF filters.
To support eBPF filters,
we add two new utilities 
  for (1) returning a file descriptor (FD) associated with an eBPF filter
  and (2) checkpointing map states by returning the FD of the $n$-th map.

Note that CRIU is only possible for privileged applications, such as container engines, as it requires \texttt{\small CAP\_SYS\_ADMIN}.
Hence, an attacker taking control of an unprivileged application cannot retrieve the FD associated with a filter and modify them.
Additionally, this mirrors the behavior of Seccomp-cBPF, where CRIU is considered secure.




\para{Container runtime integration.}
Seccomp is an essential building block for modern
  container frameworks~\cite{ncc:2016,Frazelle:2018}.
It has been mentioned that Seccomp-eBPF filters are desired~\cite{Kerner:lwn}.
Integrating Seccomp-eBPF with existing container frameworks
  is straightforward, as our implementation maintains the same
  Seccomp interface.


To demonstrate this, we have integrated Seccomp-eBPF filter support in crun~\cite{crun}, a fast OCI-compliant
  container runtime and the default container runtime of Podman~\cite{podmanio}.
The code can be found at~\cite{crun}.
Attaching an eBPF filter to a Podman container can be done with the following command:

\begin{lstlisting}[language=myJava]
podman --runtime /usr/local/bin/crun run
       --annotation run.oci.seccomp_ebpf_file
           =ebpf_filter.o
\end{lstlisting}
Contrary to cBPF filters, which are generated from JSON-based profiles
  in existing container runtimes,
  eBPF filters are directly loaded into the kernel during the
  container initialization. 
Adding such support to other container runtimes, such as runc and CRI-O, can be done in a similar way.

\para{Seccomp-eBPF filters as a privileged feature.}
For deployments that do not enable unprivileged eBPF,
  we provide a \texttt{\small sysctl} configuration
  to make Seccomp-eBPF a privileged feature.
With it set, Seccomp-eBPF can only be used by processes with the \texttt{\small CAP\_SYS\_ADMIN} capability.
Even with the configuration set, Seccomp-eBPF is still very useful for container runtimes
    and other management contexts (e.g., \texttt{\small init}) that run under the root privilege.
For example, most of the container frameworks are not rootless.

\section{Evaluation}
\label{sec:evaluation}

We evaluate the usefulness and performance of Seccomp-eBPF.
For usefulness, we use Seccomp-eBPF filters to enhance temporal
  system call specialization over
  existing cBPF filter implementations (\S\ref{sec:eval:tss}).
Moreover, we use eBPF filters to implement advanced security features
  that cannot be supported by cBPF filters.
We evaluate them
  with real-world vulnerabilities listed in Table~\ref{tab:cve}.
We then evaluate the performance of eBPF filters with both
  micro
  and macro benchmarks. We also use eBPF filters to
  accelerate stateless checks (\S\ref{sec:eval:performance}).

All experiments were run on an Intel i7-9700k with 8 cores,
  3.60 GHz, and 32GB RAM.
The machine runs Ubuntu 20.04 with Linux kernel v5.15.0-rc3, patched with our implementation of Seccomp-eBPF.



\subsection{Precise Temporal Specialization}
\label{sec:eval:tss}

\begin{figure}
  \centering
  \includegraphics[width=0.4\textwidth]{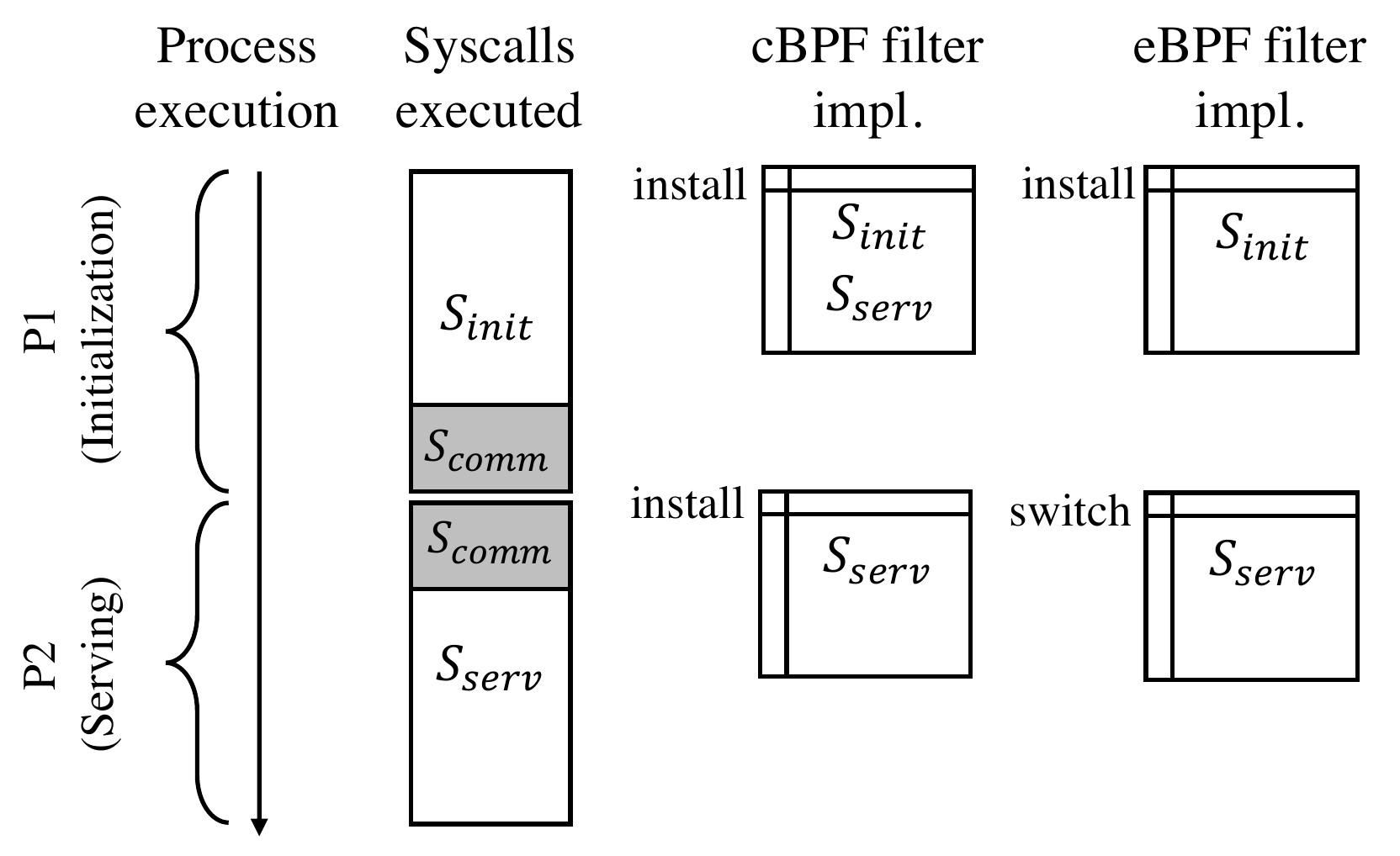}
  \caption{Two-phase temporal specialization implemented with cBPF
    and eBPF filters. $S_{init}$ and $S_{serv}$ refer to the set of
    system calls required by the initialization and serving phase, respectively;
    $S_{comm}=S_{init} \cap S_{serv}$} 
  \label{fig:tss}
  \vspace{-5pt}
\end{figure}

We explained how eBPF filters could enhance the security of temporal
  system call specialization (\S\ref{sec:usecases}).
This section quantifies the security benefits of implementing
  temporal specialization in eBPF filters and comparing them
  with existing cBPF filter implementations~\cite{Ghavamnia:usec:20}.
We evaluate temporal specialization for two distinct phases---P1 (initialization)
  and P2 (Serving), as illustrated in \cref{fig:tss}.

\para{cBPF filter.}
With cBPF filters, we install two filters:
  the first filter, installed at process startup,
  allows $S_{init} + S_{serv}$;
  the second filter only allows $S_{serv}$ and is installed
    at the program location that marks the start of the serving phase (the location
    is provided by the application developer).
Hence, cBPF filters cannot precisely implement temporal specialization, i.e., only allowing $S_{init}$ for
      the initialization phase.
As discussed in \S\ref{sec:usecases}, this problem becomes worse with
  more phases---a system call that is needed in the last phase
  has to be allowed in all the earlier phases.

\para{eBPF filters.}
We implement temporal specialization in a single eBPF filter, installed
  at application startup, in which a global variable
  is used to record the phase. The eBPF filter strictly applies different
  policies based on the phase---it allows
  only $S_{init}$ for the initialization phase and only $S_{serv}$
  for the serving phase.
Hence, it addresses the limitations of cBPF filters. The following
  snippet sketches this filter:

\begin{table}
  \footnotesize
  \centering
  \setlength{\tabcolsep}{4pt}
  \begin{tabular}{lccccc}
    \toprule
    {\bf Application} & {\bf $|S_{init}|$} & {\bf $|S_{serv}|$} & {\bf $|S_{comm}|$} & {\bf Total} & {\bf Init. Reduction}  \\
    \midrule
    HTTPD           & 71   & 83  & 47 & 107 & 33.6\% \\
    NGINX           & 52   & 93  & 36 & 109 & 52.3\% \\
    Lighttpd        & 46   & 78  & 25 & 99  & 53.5\% \\
    Memcached       & 45   & 83  & 27 & 101 & 55.4\% \\
    Redis           & 42   & 84  & 33 & 93  & 54.8\% \\
    Bind            & 75   & 113 & 53 & 135 & 44.4\% \\
    \bottomrule
  \end{tabular}
  \caption{The numbers of unique system calls in different phases of
  the evaluated applications}
  \vspace{-10pt}
  \label{tab:syscall_set}
\end{table}

\begin{table*}
  \small
  \centering
  \begin{tabular}{llll}
    \toprule
    {\bf Vulnerabilities} & {\bf Pattern} & {\bf Involved System Call(s)} & {\bf eBPF Filter as Defenses} \\
    \midrule
     \href{https://nvd.nist.gov/vuln/detail/CVE-2016-0728}{CVE-2016-0728}  & Repeated system calls & \texttt{\small keyctl} & Counter limiting \\
     \href{https://nvd.nist.gov/vuln/detail/CVE-2019-11487}{CVE-2019-11487} & Repeated system calls & \texttt{\small io\_submit} & Counter limiting \\
     \href{https://nvd.nist.gov/vuln/detail/CVE-2017-5123}{CVE-2017-5123}  & Repeated system calls & \texttt{\small waitid} & Counter limiting \\
     \href{https://bugs.busybox.net/show_bug.cgi?id=9071}{BusyBox Bug \#9071}  & System call sequences & (\texttt{\small socket} $\rightarrow$ \texttt{\small exec}) or (\texttt{\small socket} $\rightarrow$ \texttt{\small mprotect}) & Flow-integrity protection \\
     \href{https://nvd.nist.gov/vuln/detail/CVE-2018-18281}{CVE-2018-18281} & Raced system calls & \texttt{\small mremap}, \texttt{\small ftruncate} & Serialization \\
     \href{https://nvd.nist.gov/vuln/detail/CVE-2016-5195}{CVE-2016-5195}  & Raced system calls & (\texttt{\small write}, \texttt{\small madvice}) or (\texttt{\small ptrace}, \texttt{\small madvice}) & Serialization \\
     \href{https://nvd.nist.gov/vuln/detail/CVE-2017-7533}{CVE-2017-7533}  & Raced system calls & \texttt{\small fsnotify}, \texttt{\small rename} & Serialization \\
    \bottomrule
  \end{tabular}
  \caption{Evaluated vulnerabilities and their patterns that can be effectively defended by Seccomp-eBPF filters.}
  \vspace{-10pt}
  \label{tab:cve}
\end{table*}

\begin{lstlisting}[language=myJava,label={lst:tss_ebpf}]
unsigned int phase;
SEC("seccomp")
int bpf_prog_ts(struct seccomp_data *ctx) {
  if (!phase && ctx->nr == -1) {
    phase = 1; // phase change
    return SECCOMP_RET_ALLOW;
  }
  if (!phase) { // only allow S_init
    switch (ctx->nr) {
      case SYS_alarm:
        return SECCOMP_RET_ALLOW;
      ...
      default: break;
    }
  } else { // only allow S_serv
    switch (ctx->nr) {
      case SYS_accept:
        return SECCOMP_RET_ALLOW;
      ...
      default: break;
    }
  }
  return SECCOMP_RET_ERRNO | EPERM;
}
\end{lstlisting}

Different from cBPF filters, where the second filter needs to be installed
  at the phase-changing location,
  we insert a dummy system call that marks the phase change.

\para{Attack surface reduction.}
We use the six server applications in
  the temporal specialization work~\cite{Ghavamnia:usec:20}.
We use $S_{init}$ and $S_{serv}$ of each application,
  provided by the research artifact~\cite{Ghavamnia:usec:20},
  from which we derive $S_{comm}$.
\cref{tab:syscall_set} shows the attack surface reduction of the initialization
  phase achieved by eBPF filters over the cBPF filters.
The eBPF filters reduce the attack surface of the initialization phase
  by 33.6\%--55.4\% across the evaluated applications.


\subsection{New Security Features}
\label{sec:eval_new_features}

We present new security features enabled
  by Seccomp-eBPF filters
and show that they can effectively mitigate real-world vulnerabilities (Table~\ref{tab:cve}).

\subsubsection{Count and rate limiting}
\label{sec:eval:count}

We implement eBPF filters for count limiting
  which only allows
  a system call to be invoked a limited number of times (\S\ref{sec:usecases}).
The filter keeps the count of the target system call using a global variable and
  increments the count when the system call is invoked.
Once the count reaches a predefined value $N$, all subsequent invocations
  of the specific system call are denied.

The following code shows an eBPF filter that only allows an application
  to call the \texttt{\small keyctl} system call with
  the argument \texttt{\small KEYCTL\_JOIN\_SESSION\_KEYRING}
  up to a max allowed count---most applications
    (e.g., \texttt{\small e4crypt}) only need to call \texttt{\small keyctl} once or twice.
\begin{lstlisting}[language=myJava,label={lst:keyctl}]
unsigned int count;
SEC("seccomp")
int bpf_prog_cnt(struct seccomp_data *ctx) {
  if (ctx->nr == SYS_keyctl &&
      ctx->args[0] == KEYCTL_JOIN_SESSION_KEYRING) {
      if (count >= MAX_ALLOWED_CNT)
         return SECCOMP_RET_ERRNO | EPERM;
      count += 1;
  }
  return SECCOMP_RET_ALLOW;
}
\end{lstlisting}

The eBPF filter can mitigate vulnerabilities
  like CVE-2016-0728 which is exploited
  by repeatedly calling \texttt{\small keyctl}
  with \texttt{\small KEYCTL\_JOIN\_SESSION\_KEYRING} to create invalid keyring objects.
This would trigger buggy error handling code that omits to decrement
  the refcount and thus cause the refcount to overflow.
We also evaluate count limiting with
  CVE-2019-11487 and CVE-2017-5123 shown in \cref{tab:cve}.

Also, we implement {\it rate} limiting
  for frequency of specific system calls,
  using timer helpers (\cref{fig:helpers}).

\subsubsection{Flow integrity protection}
We use Seccomp-eBPF filters to implement the kernel enforcement of system
  call flow-integrity protection (SFIP)~\cite{Canella:2022} which otherwise
  requires kernel revisions and a new system call.
SFIP checks both system call sequences and origins (code address that issues
  a system call).
For the sequence check,
  we store the system call state machine in a two-level eBPF map,
  by encoding the state machine as a $N \times N$ matrix where
  $N$ is the number of system calls an application uses.
We maintain the previous system call ID $s'$ in a global variable.
Given a system call $s$, we check whether $s' \rightarrow s$ is valid.
For the origin check, we store the system call origin mapping
  in another two-level map
  where the first-level map indexes the system call ID and the second-level stores the valid code addresses.
For a system call $s$, we check whether $s$'s origin is included in the
  second-level map after indexing the first-level map with $s$'s ID
(Seccomp already provides the calling address of a system call in its data structure).

We evaluate Seccomp-eBPF filter-based SFIP using the same buffer overflow vulnerability
  in BusyBox~\cite{busybox:bug:9071} in the evaluation of the original
  SFIP work~\cite{Canella:2022}. The exploits require a system call sequence
  of either \texttt{\small socket} $\rightarrow$ \texttt{\small execve} to execute shellcode, or
  \texttt{\small socket} $\rightarrow$ \texttt{\small mprotect} to mark the memory protection
  of the shellcode as executable.
Neither of the transitions is allowed by the eBPF filter based on the in-map
  system call state machines. Furthermore, the code addresses are checked
  as per the origin map.

\subsubsection{Serialization.}

\new{
We implement the eBPF based serializer (see \S\ref{sec:syscall_serialize}) to
  mitigate the race-based vulnerabilities in \cref{tab:cve}.
We write applications that issue the system calls of
  the race vulnerabilities concurrently.
Take CVE-2018-18281 as an example,
  the filter serializes
  \texttt{\small mremap} and \texttt{\small ftruncate} when the applications issue both.
}

\subsection{Performance}
\label{sec:eval:performance}

We compare the performance of eBPF filters
  with cBPF filters and Notifier that implement the same security policy.
For fair comparison, we use policies that can be implemented by cBPF filters.
We expect the performance of eBPF filters and cBPF filters to be similar,
  since Seccomp internally converts cBPF filters into eBPF code.
On the other hand, eBPF filters and cBPF filters go through different toolchains
  and thus are optimized differently.


\subsubsection{Microbenchmark}
\label{sec:eval:microbenchmark}

\cref{tab:microbench} shows our microbenchmark results of different
  Seccomp filters, including the
  execution time (cycles) of the \texttt{\small getppid} system call
  and the filter programs.
We use a policy that denies 245 system calls and allows the rest (\texttt{\small getppid} is allowed).
To obtain reliable results, we fix the CPU frequency and disable Turbo boost.

We generate the two cBPF filters using libseccomp (v2.5.2), with the default
  option and with binary-tree optimization~\cite{Hromatka:2018:2}.
The latter generates a filter that sorts the system call ID
   in a binary tree.
We implement the eBPF filter in C and use Clang (\texttt{\small -O2}) to compile
  the C code into eBPF bytecode.
Clang optimizes switch-case statements into a binary tree so that reaching
  each case only takes $O(log(N))$ time, where $N$ is the number of system calls.
We also experiment with Seccomp constant-action bitmap~\cite{bitmap}, an
  optimization that
  skips the filter execution if the system call ID is known to be allowed.
For Seccomp Notifier, we implement a userspace agent in C with the same logic as the eBPF filter.

Our results show that the eBPF filter outperforms the unoptimized cBPF filter.
It has roughly the same performance as
  the cBPF filter optimized by libseccomp.
Both Clang and libseccomp optimize the filter code using binary search
  to avoid walking over a long jump list.
Constant-action bitmap achieves the highest performance, but it cannot
  help policies on argument values.
With Seccomp Notifier, \texttt{\small getppid} runs 45.4 times slower than with
  the eBPF filter.

\begin{table} 
  \footnotesize
  \centering
  \begin{tabular}{lcc}
    \toprule
    {\bf } & {\bf getppid (cycles)} & {\bf filter (cycles)} \\
    \midrule
    No filter           & 244.18 & 0 \\
    cBPF filter (default)       & 493.06 & 214.19 \\
    cBPF filter (optimized) & 329.47 & 68.68 \\
    eBPF filter              & 331.73 & 60.18 \\
    Constant-action bitmap~\cite{bitmap}     & 297.60 & 0 \\
    Seccomp notifier     & 15045.05 & 59.29 \\
    \bottomrule
  \end{tabular}
  \caption{Execution time of system calls with different types of
    filters and filter execution time.}
    \vspace{-10pt}
  \label{tab:microbench}
\end{table}

\subsubsection{Application Performance}
\label{sec:eval:tss:perf}

\begin{table}
  \footnotesize
  \centering
  \begin{tabular}{lll}
    \toprule
    {\bf Application} & {\bf Description} & {\bf Benchmark} \\
    \midrule
    HTTPD           & Web server & ab~\cite{ab} (40 clients) \\
    NGINX           & Web server & ab~\cite{ab}  (40 clients) \\
    Lighttpd        & Web server & ab~\cite{ab} (40 clients) \\
    Memcached       & In-memory cache & memtier~\cite{memtier}  (10 threads) \\
    Redis           & Key-value store & memtier~\cite{memtier}  (10 threads) \\
    Bind            & DNS server & dnseval~\cite{dnseval}  \\
    \bottomrule
  \end{tabular}
  \caption{Applications and benchmarks used in evaluation}
  \vspace{-10pt}
  \label{tab:app}
\end{table}

We measure the application performance with different types of Seccomp filters.
We use temporal specialization as the security policy.
The implementations of the cBPF filters and eBPF filters are explained in \S\ref{sec:eval:tss}.
Seccomp Notifier can also support precise temporal specialization like
  eBPF filters.
We implement a version that defers all decisions
  on system calls to a user agent.
The handler keeps a phase-changing flag, which is set when the application
  enters the serving phase.
We also evaluate an optimization that combines cBPF filter and
  the agent, denoted as {\it hybrid}.
The hybrid version installs a cBPF filter at the process startup,
  allowing $S_{init}$.
For every system call outside of $S_{init}$, the filter defers the decision to the user space.
An additional filter is installed at the beginning of the serving phase to block $S_{init}-S_{comm}$.
The handler then only allows $S_{serv}-S_{comm}$ after the phase change.



\begin{figure}
  \centering
  \includegraphics[width=0.45\textwidth]{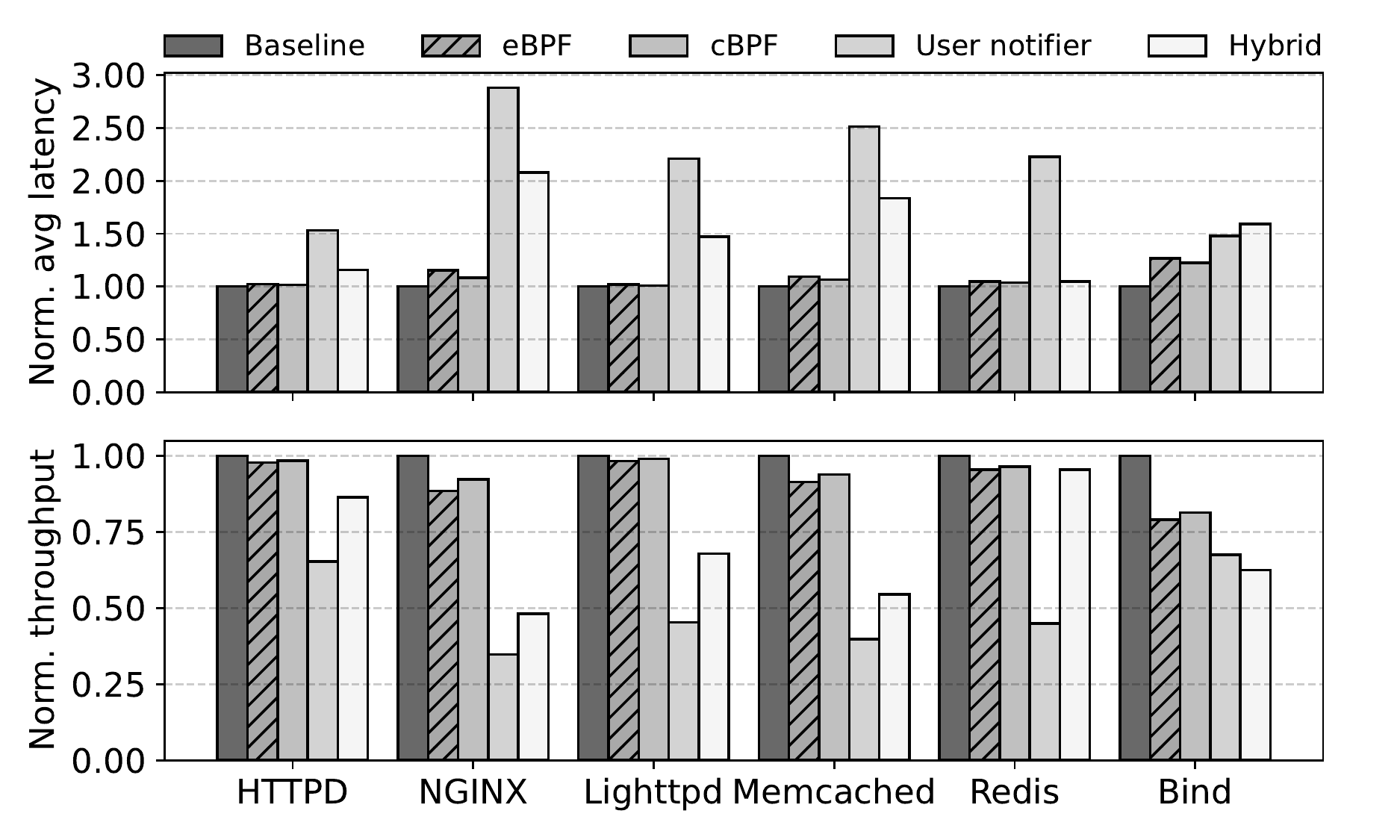}
  \vspace{-5pt}
  \caption{Avg. latency and throughput of the applications with
     different filters that implement temporal specialization
     (normalized to the baseline that disables Seccomp).}
  \label{fig:eval:tssperformance}
  \vspace{-5pt}
\end{figure}

\para{Applications and benchmarks.} We use the same six server applications
  as in \S\ref{sec:eval:tss}.
We use official benchmark tools for each
  application (\cref{tab:app}).
All experiments are run 10 times and the average numbers are reported.
\cref{fig:eval:tssperformance} shows the average latency and throughput
  of each application with
  cBPF, eBPF, Notifier, and Hybrid.
We normalized all results to the \textit{baseline}, a version that does not use Seccomp.
The results show two consistent characteristics.
First, Seccomp-eBPF has similar
  performance impacts as Seccomp-cBPF, while
  providing higher security for the initialization phase.
Second, userspace agents incur significant performance overhead.
Applications with pure user notifiers have additional
48\%--188\% average latency and 32\%--65\% lower throughput.
Even for Hybrid, there is an additional 5\%--108\% average latency and
5\%--52\% lower throughput.
NGINX suffers from the highest degradation, with
  latency increasing by a factor of 1.88 and throughput decreasing
  by 65\%.
Only Hybrid for Redis has a performance comparable to BPF filters,
  because in
  the serving phase of Redis, most system calls are from
  $S_{comm}$, which are filtered by cBPF filters.

\subsubsection{Accelerating stateless security checks}

Lastly, we use eBPF filters to implement the Draco Seccomp cache~\cite{skarlatos:micro:20}
  and repeat the Draco evaluation using the applications and benchmarks
  in Table~\ref{tab:app}.
The key idea of Draco is to cache the ID and the corresponding argument values
  of a system call that has recently been validated by stateless security
  checks. If an incoming system call hits the cache,
    Draco saves the computation of running the stateless checks.
Draco is effective when the stateless checks are expensive and the system calls
  of an application have high locality.
The eBPF filter implements Draco using an array map which
  maps a system call to its corresponding check filter through an eBPF tail-call.
The check filter uses a hash map to store recently validated argument
  values in a 48-byte blob.

We use the profile which only allows
  a system call if its ID and argument values are recorded by \texttt{\small strace}
  in a dry run. Our results show that eBPF-based Draco increases the
  throughput of three web servers by 10\% on average.
We do not observe significant improvement for the other applications.

\section{Discussion}
\label{sec:discussion}

{\bf System call filtering with LSM.}
\new{Linux Security Modules (LSMs) provide enforcement
  for system-wide security policies.
Therefore, it is commonly discussed together with Seccomp,
  often as an alternative to implement system call filtering,
  especially with LSM-eBPF~\cite{Edge:krsi}.
In fact, Seccomp(-eBPF) and LSM(-eBPF) are fundamentally different.}



\new{Seccomp restricts the system call entry point,
    while LSMs provide access control on kernel objects deeply on the path of
    system call handling.
  Hence, LSMs cannot prevent vulnerabilities before reaching LSM hooks.
  Moreover, Seccomp can provide fine-grained
    per-process system call filtering,
    while LSMs are system-wide.
Lastly,  LSM requires privileged use cases (only sysadmins
    can apply LSM policies);
Seccomp-eBPF supports
    unprivileged use cases where an unprivileged process
    can install a Seccomp-eBPF filter.
}

\new{Seccomp and LSM have different principles.
Seccomp requires no new kernel code; hence, it
  stays future-proof for new sysem calls;
LSMs would need new code when new system calls are added.
So, code in a Seccomp-eBPF filter takes a ``microkernel'' style.
LSM uses a ``monolithic kernel'' style to
  implement custom checks for different system calls.}


\para{Automatic Seccomp-eBPF filter generation.}
Many tools have been developed to {\it automatically} generate
  application-specific
  Seccomp-cBPF filters.
The basic idea is to profile the system calls of the target
  application with static code analysis~\cite{Ghavamnia:raid:20,Ghavamnia:usec:20,Pailoor:oopsla:20,Canella2021chestnut},
  binary analysis~\cite{Canella2021chestnut,DeMarinis:20}, or dynamic tracking~\cite{Wan:2017,Rothberg:podman:gen}.
The identified system calls are included in a static allow list.


How to automatically generate Seccomp-eBPF filters
  is an open question.
Recent work like SFIP~\cite{Canella:2022} shows promises of
  generating system call state machines.
We believe that eBPF filters for count/rate limiting and temporal
  specialization can also be automatically generated.
Moreover, many ideas from system call based
  intrusion detection using data modeling and machine learning
  can potentially be applied
  to generate advanced eBPF filters~\cite{Mutz:2006,Maggi:2010,Linn:2005,Warrender:1999,Forrest:1996,Feng:2003,Ghosh:1999}.

\para{Risks of unprivileged eBPF.}
With bugs in Linux
  eBPF verifier and JIT compiler~\cite{Nelson:2020},
  Seccomp-eBPF may potentially allow
  unprivileged attackers to craft malicious eBPF filters to exploit vulnerabilities.
Seccomp-cBPF shares the same risk; however, since it is simpler it likely has fewer bugs.
We believe that in the long term unprivileged eBPF will be safe,
  as evidenced by
recent work on formally-verified eBPF
  verifiers and compilers~\cite{Xu:2021,Nelson:2020,Geffen:2020,Wang:2014,Gershuni:pldi:2019},
  and more broadly, safe and correct kernel code~\cite{Nelson:sosp:19,Elhage:2020,Li:hotos:2021,Balasubramanian:hotos:2017}.

\comments{
In fact, unprivileged eBPF is already used
  (e.g., in socket filters and cgroup socket buffers \cite{bpf:privcheck:kernel}),
  no matter whether
  Seccomp supports eBPF filters or not.
}

Note that Seccomp-eBPF can be configured as a privileged feature at deployment
  (\S\ref{sec:impl}),
  if unprivileged eBPF is a concern.
The privileged configuration can be used by many container runtimes and
  management services (e.g., \texttt{\small init}).

\section{Related Work}
\label{sec:related_work}

{\bf System call filtering.}
Prior work has studied system call filtering (aka interposition) techniques
    for
    protecting the shared OS kernel against untrusted applications~\cite{Goldberg:1996,
Wagner:1999,Provos:2003,Jain:2000,Garfinkel:2004,Acharya:2000,Kim:2013,
Alexandrov:1999,Peterson:2002,Fraser:2000,Linn:2005,Dan:1997,Ghormley:1998,Douceur:2008}.
Early techniques rely on userspace agents (e.g., based on \texttt{ptrace})
    that check user-specified
    policies to decide which system calls to allow or deny,
    in the same vein as Seccomp Notifier (see \S\ref{sec:background:notifier}).
However, 
    the context switch overhead could be unaffordable to applications
    that require high performance.
Seccomp provides a solution that allows user-specified policies to
    be implemented in cBPF and executed as kernel
    extensions. 
Compared with userspace agents, Seccomp filters have significant performance
    advantages, which is one main reason that makes it a widely-used
    building block for modern sandbox and container technologies.
Our work builds on the success of Seccomp and rethinks
    the programmability of system call security in the Seccomp context.





\para{Discussions on eBPF Seccomp filters in the Linux community.}
There were a few other proposals on supporting eBPF filters for Seccomp, with patches~\cite{Dhillon:2018,Hromatka:2018}.
However, our discussions with the community both on the kernel mailing list~\cite{Zhu:2021}
    and at the Linux Plumbers Conference~\cite{Jia:lpc:2022}
    tell that it is still very controversial.

One common concern is lacking concrete use cases,
    as exemplified by the maintainer's response---``{\it What's the
    reason for adding eBPF support? Seccomp shouldn't need it...
    I'd rather stick with cBPF
    until we have an overwhelmingly good reason to use eBPF...}''~\cite{kees:use_case}.
One reason is that early patches~\cite{Dhillon:2018,Hromatka:2018} do not support
    maps and thus still have no statefulness.
We hope that our work addresses this concern.
Our design is driven by an analysis of desired system call filtering features,
    which reveals the limitations of Seccomp (\S\ref{sec:usecases}).
We also show that existing eBPF utilities are insufficient;
    therefore, simply opening the eBPF interface cannot solve the problem.




There are other concerns on eBPF security, including
    (1) exposure of kernel data and functions to untrusted user space
    via maps and helpers,
    and (2) potential vulnerabilities in the eBPF subsystem.
In our design, the security of Seccomp-eBPF can be systematically
    reduced to that of Seccomp and eBPF.
We discuss the risk of unprivileged eBPF in \S\ref{sec:discussion}.
We believe that vulnerabilities in the eBPF subsystem
    implementation
    is a temporal (but challenging) problem that will be addressed in the longer term.
Our implementation also provides a configuration option
    to turn Seccomp-eBPF into a root-only feature (\S\ref{sec:impl})
    which is useful for container environments.

\para{Other eBPF use cases.}
Recently, eBPF has been actively used
    to build innovative tools and systems,
    ranging from networking~\cite{Hoiland-Jorgensen:conext:2018}
    to tracing~\cite{bpf_tracing} to storage~\cite{Zhong:osdi:2022,BMC}
    to virtualization~\cite{Amit:atc:2018}.
Different from most of the eBPF use cases,
    Seccomp needs to support
    unprivileged use cases (\S\ref{sec:threat_model}).
Therefore, security is a first-class design principle of Seccomp-eBPF.

\section{Concluding Remarks}
\label{sec:conclusion}

Our work makes one step towards empowering advanced system call security policies
    by enhancing the programmability of system call security mechanisms, namely Seccomp in Linux.
We present our design and implementation of the Seccomp-eBPF program type
    and show that it can enable many useful features,
    without impairing system call performance or
    reducing system security.
Our work is imperfect, but we hope that it could lay the foundation 
    and motivate
    strong programmability for system call security.

\section*{Acknowledgement}
This work was supported in part by the IBM-Illinois Discovery
Accelerator Institute and by NSF grant CNS-1956007.
We thank Giuseppe Scrivano for helping with the eBPF Seccomp filter 
    support in crun.
We thank Kele Huang, Yicheng Lu, Austin Kuo, and Josep Torrellas for early participation of the work.
We thank Michael Le, Mimi Zohar, and Hani Jamjoom for the discussions of the work.
We also thank developers from the Linux community for discussing the work 
    with us on the Linux kernel mailing list and at LPC 2023.

\bibliographystyle{plain}
\bibliography{ref,draco}

\end{document}